\documentclass[a4paper,10pt]{article}
\AtBeginDocument{%
  \providecommand\BibTeX{{%
    \normalfont B\kern-0.5em{\scshape i\kern-0.25em b}\kern-0.8em\TeX}}}

\usepackage{graphicx}

\usepackage{subcaption} 
\usepackage{listings}

\usepackage{float}

\usepackage[ruled]{algorithm2e} 
\usepackage[utf8]{inputenc}

\SetAlFnt{\small}
\SetAlCapFnt{\small}
\SetAlCapNameFnt{\small}
\SetAlCapHSkip{0pt}
\IncMargin{-\parindent}

\newcommand{\SYSMARK}{SYSmark}
\newcommand{\SYSMARKtwe}{SYSmark~2012}
\newcommand{\SPECCPU}{SPEC~CPU}
\newcommand{\SPECCPUsix}{SPEC~CPU~2006}

\begin{document}

\title{Employing Simulation to Facilitate the Design of Dynamic Code Generators}

\author{Vanderson M. do Rosario$^1$ \and Raphael Zinsly$^{1,2}$ \and Sandro Rigo$^1$ \and Edson Borin$^1$}
\date{%
    $^1$Institute of Computing - UNICAMP - Brazil\\%
    $^2$IBM - Campinas - Brazil\\[2ex]%
    \today
}

\maketitle

\begin{abstract}
  Dynamic Translation (DT) is a sophisticated technique that allows the implementation of high-performance emulators and high-level-language virtual machines.
  In this technique, the guest code is compiled dynamically at runtime. Consequently, achieving good performance depends on several design decisions, including the shape of the regions of code being translated. Researchers and engineers explore these decisions to bring the best performance possible. However, a real DT engine is a very sophisticated piece of software, and modifying one is a hard and demanding task. 
  Hence, we propose using simulation to evaluate the impact of design decisions on dynamic translators and present RAIn, an open-source DT simulator that facilitates the test of DT's design decisions, such as Region Formation Techniques (RFTs). 
  RAIn outputs several statistics that support the analysis of how design decisions may affect the behavior and the performance of a real DT.
  We validated RAIn running a set of experiments with six well known RFTs (NET, MRET2, LEI, NETPlus, NET-R, and NETPlus-e-r) and showed that it can reproduce well-known results from the literature without the effort of implementing them on a real and complex dynamic translator engine.
\end{abstract}

\section{Introduction}


Emulators and high-level-language virtual machines compile applications' code during their execution.
This approach, known as Just-In-Time (JIT) compilation or Dynamic Translation (DT), is a concept that is as old as high-level programming languages themselves~\cite{aycock03}.
DT techniques can be used to improve execution time and space efficiency of programs~\cite{brown76} and to support programming language versatility with High-Level Language Virtual Machines (HLLVM)~\cite{adams14}. 
They can also be used to maintain support for legacy code by the industry~\cite{dehnert03,apple} or to support new architectures such as RISC-V \cite{lupori2018towards,do2018fog}.
In dynamic high-level languages, DT can be used to emulate their intermediate representation, such as the Facebook Hip-Hop Virtual Machine~\cite{adams14}, the Java HotSpot~\cite{suganuma00}, and Firefox's IonMonkey JavaScript JIT~\cite{IonMonkey17}.

In order to achieve high-performance using dynamic translation, the cost added by invoking a compiler during runtime should be lower than the performance gains achieved by the execution of the code produced by the compiler. 
To pay off the compilation cost, a piece of code needs to execute for a significant amount of time, so the savings achieved by the execution of the optimized code are also significant. 
Fortunately, programs usually spend most of their execution time in a minority of their code~\cite{knuth71}, and DT can achieve high-performance by only translating and optimizing frequently executed code~\cite{cad13}, which we call hot code. 
For the rest of the application, the cold part, an interpreter, or a fast-non-optimizer compiler can be used.
Therefore, one of the main strategies to improve the performance of a DT engine (DTE) is to create heuristics to predict, at runtime, which part of the code is hot and which is cold. 
In this paper, the term Region Formation Techniques (RFTs) will be used in its broadest sense to refer to all these prediction schemes. 
The quality of the code produced by the dynamic translator for hot regions is also an essential factor that affects performance. 
In this context, both the set of optimizations employed and the granularity and shape of the region of code being compiled play an important role in the code quality and, hence, its performance.
For instance, while small portions of code, such as basic blocks, can be fast compiled, larger ones expose more opportunities for optimizations~\cite{suganuma06}. 
RFTs that capture whole loops or even code from more than one method in the same region enables more aggressive (loop or inter-procedural) optimizations. Moreover, besides affecting the scope of optimizations, the RFT design decisions may also affect the hot code frequency, the code duplication rate, and the optimization costs, among others. 

All these variables need to be considered while designing and evaluating a DTE, and this is not a trivial task. 
A DTE is a sophisticated piece of software that includes in itself a compiler, may also include an interpreter, complex data structures for storing compiled binary, a linker, and code to orchestrate all these pieces together. Understanding the code of a DTE or debugging one is challenging, mainly when the bug occurs on the code generated dynamically by the DTE. 
Consequently, in this kind of software, implementing and validating novel research and design ideas is usually a complex process.
In fields like processor architecture designing, prototyping ideas in real hardware is also complex and simulation is broadly used to make design exploration approachable~\cite{burger1996evaluating, 10.1007/3-540-63875-X_40,998259,6893220}. 
In this work, we argue that simulation can also be used to facilitate DT design exploration, and we present RAIn, a novel DT/RFT simulator.  


We use RAIn to evaluate several RFTs and reproduce results from the literature allowing its simulation capabilities validation. 
Our evaluation setup includes programs from both the SPEC-CPU 2006~\cite{SPEC2006} and SYSMARK~\cite{Sysmark} benchmarks and covers six different RFTs techniques from the literature. The contributions of this paper can be summarized as follows:

\begin{itemize}
    
    \item A \textbf{novel DTE Simulator}, called RAIn, which makes the implementation and testing new RFTs simpler and faster. 
    RAIn is capable of producing several different statistics that facilitate the evaluation of the behavior and performance of the RFTs. 
    
    \item A \textbf{comprehensive study} of the performance and behavior of six RFTs (NET~\cite{Net}, MRET2~\cite{Mret2}, LEI~\cite{Lei}, Netplus~\cite{Netplus}, Relaxed NET~\cite{hong2014}, and Extended Netplus~\cite{hsu2015}) using programs from SPEC-CPU 2006~\cite{SPEC2006} and SYSMARK~\cite{Sysmark}, thus covering several different application profiles. 
    The results corroborate the findings encountered by previous work and provides a comparison of all the techniques using the same set of applications.

\end{itemize}

The remainder of the text is organized as follows:
Section~\ref{sec:trace-based} discusses the typical organization of a trace/region-based DTE and the characterization of their overhead (performance issues).
Section~\ref{sec:rain} describes the proposed simulator, including its functionality and its advantages.
Section~\ref{sec:setup} shows the experimental setup, and Section~\ref{sec:results} presents a comprehensive comparison between several DT designs.
Finally, Section~\ref{sec:conclusion} presents the conclusions.
\section{Region-based Dynamic Translators}
\label{sec:trace-based}

A Dynamic Translator Engine (DTE) is a piece of software that translates guest code, which may be a binary generated for one computer architecture, into code compatible with the host architecture, also known as native code. 
When emulating hot code, i.e., frequently executed code, it usually pays off to spend effort translating and optimizing the guest code into optimized native code. In this case, the performance gains achieved by executing the optimized native code surpasses the costs of translating the code.
For cold code, i.e., infrequently executed code, it is usually better to employ techniques such as interpretation or quick, basic-block-based, translation, which have no or low translation cost.
In this paper, we use the term interpretation to represent the mechanisms employed to emulate cold code. 

\begin{figure}[H]
	\centering
	\includegraphics[width=0.6\columnwidth]{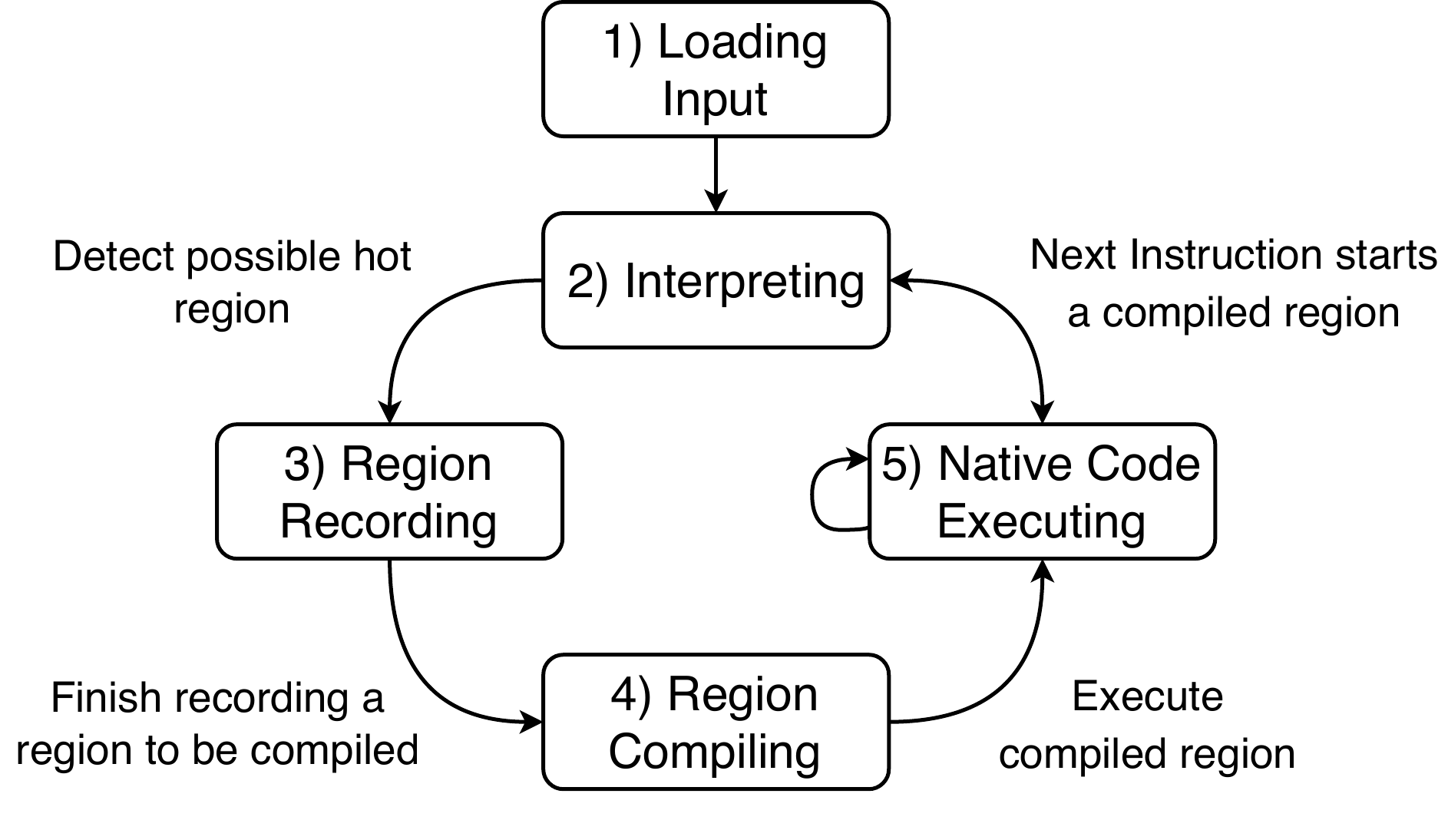}
	\caption{\label{jitlifetimeFig}Execution flow of a common DTE.}
\end{figure}

Figure~\ref{jitlifetimeFig} illustrates the execution flow of a common DTE. 
First, it loads the guest code to memory (state 1), which can be, for example, an intermediate representation such as Java Byte Code or an x86 binary. 
Then, the emulation process begins by fetching, decoding, and interpreting all the instructions one by one (state 2), in a process called interpretation. 

During the interpretation, an active monitor (profiler), or the interpreter itself, monitors whether the emulation is repeatedly executing the same code for longer than a given threshold, called hotness threshold. 
If so, this means that the execution is on a hot part of the code, such as a cycle. 
Once detected that the execution is in a hot code, the interpreter starts to record the trace of instructions to form a region of code for translation (state 3). 
These recorded instructions, normally part of a loop or a cycle in the static code, are passed to a compiler which compiles the region from the guest architecture into a semantically equivalent optimized code compatible with the host architecture, the native code (state 4). 
The native code is then stored in a code cache and every time this same piece of code needs to be executed in the future, the execution jumps to the native code in the cache (state 5) instead of interpreting it.
All these steps are repeated until the entire program's execution is finalized.

DTEs implement specific functionalities to execute each one of the aforementioned steps. For example, to predict which regions of code are hot or not, there are three typical implementations~\cite{bond2005practical}: a) frequency counting based on instrumentation, b) sampling based on interrupt-timers, or c) a combination of both. 
To store the compiled code and make it fast to access, DTEs employ hash maps organized as caches, called Translated Code Cache, or TCC. 
Another critical process during emulation is mapping guest addresses into their respective emitted host code addresses.
As translation does not always result in the same memory layout, access to memory locations using the address from the guest code in indirect jumps and returns needs to be mapped to the address in the compiled/translated host code. 
The mechanism that handles this mapping of addresses during region execution is usually referred to as Indirect Branch Translation Handler~\cite{hiser07}.
All these mechanisms and structures carry design decisions and details that directly affects the performance of a DTE.

Another important mechanism is the Region Formation Technique (RFT), which is responsible for :
(1) deciding which instructions to profile,
(2) deciding when to start recording regions, and
(3) deciding when to stop recording them. 
At one hand, to minimize compilation overhead, it is important to translate only hot code. 
On the other hand, to accelerate the native code, it is usually necessary to form large regions to increase the translation scope and expose more optimization opportunities to the optimizer. 
For example, RFTs that capture whole loops or even code from more than one method in the same region enables more aggressive (loop or inter-procedural) optimizations. 
Consequently, the RFT design may have a big impact on the performance of the native code and, hence, the DTE.


So far, the main RFTs proposed in the literature  are NET (a.k.a MRET)~\cite{Net}, NET-r~\cite{hong2012hqemu}, MRET2~\cite{Mret2}, LEI~\cite{Lei}, NETPlus~\cite{Netplus} and NETPlues-e-r~\cite{guan2010distribit}. 
All of them using different strategies to select hot code and selecting dynamic regions with different sizes, shape, and characteristics which directly affects the performance of a DTE. 
Below, we include a brief description of each one of them: 

\begin{itemize}

\item \textbf{NET}: The authors of Dynamo~\cite{Dynamo} introduced an RFT called NET (Next-Executing Tail)~\cite{Net}, which was originally called MRET (Most Recent Executed Tail).
In NET, regions are superblocks.
Targets of backward branches or targets of other superblock exits are considered as potential superblock entries and are assigned a counter that keeps track of its execution frequency.
After the counter reaches a defined hotness threshold, a new region is recorded starting from this instruction and continuing until another backward branch is reached or a given maximal number of instructions is included in the region.

\item \textbf{MRET2}: The authors of StarDBT~\cite{Stardbt} introduced the MRET2 RFT, a variation of NET that aims at reducing the number of side-exits~\cite{Mret2}.
MRET2 consists of executing the recording phase of the NET technique twice. 
If different code sequences are selected during both recordings, only the intersection between them is selected to compose the MRET2 region.

\item \textbf{LEI}: Hiniker, Hazelwood, and Smith~\cite{Lei} introduced the {\it Last Executed Iteration} (LEI) technique.
LEI selects cyclic superblocks based on a history buffer for the current execution.
It focuses on avoiding inner loop duplication on the superblocks.

\item \textbf{NETPlus}: Davis and Hazelwood pointed out in a more recent work~\cite{Netplus} that the history buffer used by LEI imposes a considerable overhead.
In this latter paper, the authors propose the NETPlus RFT, which follows the same steps as NET up to the point where a region is being closed.
At this point, NETPlus will look ahead in the code for a branch whose target is the beginning of the superblock.
When found, all instructions between the current end of the region and the branch are added to the superblock.
In this manner, NETPlus aims at capturing more loops inside individual superblocks when compared to the original NET, imposing a low overhead on the superblock selection process.
It is important to notice, however, that the look-ahead process may touch code or memory positions that have never been touched before and may never be touched in the future, which may trigger unexpected page faults.
The deepness in which the search can go is a parameter for the NETPlus RFT.

\item \textbf{NET-r and NETPlus-e-r}: Hong et al.~\cite{hong2012hqemu} presented a modified version of QEMU that uses the LLVM backend to emit highly optimized regions, named HQEMU.
The authors observe that to obtain maximum benefits from the LLVM optimizations, HQEMU needs to create large regions of code.
Thus, they present a modification of two known RFTs.
The first, called NET-r, is a relaxation of NET that makes it similar to the cyclical-path-based repetition detection scheme by not end recording a region when a backward branch is found, but when a cycle is found (repeated instruction address recorded). 
The second~\cite{guan2010distribit}, called NETPlus-e, is an extension of NETPlus that adds not only paths that exit the NET region and returns to its entrance but also paths that exit the NET region and returns to any part of the region. 
NETPlus-e can also use the NET-r instead of NET, thus creating an extended and relaxed version of NETPlus (NETPlus-e-r). 
\end{itemize}

\subsection{Dynamic Translation Performance}
\label{sec:overheads}

In this section, we discuss the primary sources of overhead in a DTE, mainly the ones related to the RFT choice. 

If we consider emulation flow depicted in Figure~\ref{jitlifetimeFig}, at any given time, a DTE can be in any of the five states.
State 1, Loading DTE and Guest Code, only needs to be executed once, and for long-time executions, it incurs  a minimum overhead; hence, we will not consider it. 
Instead, we will focus on the performance and overhead sources on the other four: Interpretation, Region Record, Region Compilation, and Native Execution.


The total execution cost of emulating code with a DTE is composed of the cost of interpreting (State 2) and profiling (State 3) cold code plus the cost of compiling hot code (State 4) and the cost of executing the native code (State 5). 
Notice that the sooner a code is compiled, the lesser time the DTE spends emulating it as cold code and more time it spends emulating it as hot code, i.e., executing optimized native code.


Emulating code with native/optimized code is faster than emulating code with interpretation, so one greed approach would be to compile every single part of the code, but we need also to consider the compilation cost. If the execution frequency is low, the compilation overhead may exceed the gains achieved by the hot-code emulation. In the case of cold code, compiling damages the final performance instead of improving it.
In this case, to only interpret the code is the best option~\cite{cad13}.

This can be summarized by equations~\ref{eq:1}, \ref{eq:2} and \ref{eq:3}, where $Interp_{Cost}$ is the average cost of interpreting each guest instruction, $InterpFreq$ is the number of instructions interpreted, $HotStaticSize$ is the total number of guest instructions dynamically compiled, $Gen_{cost}$ is the average cost of compiling a single guest instruction, $CompilerInitializationCost$ is the initialization overhead of calling the compiler, $NumRegions$ is the number of regions of code being compiled, $Native_{Cost}$ is the average cost of emulating a guest instruction by executing native, compiled, code, $NativeFreq$ is the number of guest instructions emulated by native code, $TotalFreq$ is the total number of guest instructions emulated.
Notice that compiling a code only results in performance gains when the inequality of Equation~\ref{eq:4} is true. 

Other important performance overhead in a DTE is the region transition overhead. Transitioning between the interpreter and a native region of code or transitioning between native regions of code may imply in saving and loading emulation context. 
Emulators may maintain a context of the machine being emulated, such as the values of the registers. 
In native code, these guest registers can be mapped to host registers, but when jumping to the interpreter these values need to be saved to memory so it can be again accessed by the interpreter. 
The same happens when regions are compiled with a register allocation that chooses different guest-host register mapping per region. 
In this case, the guest registers modified that are in host registers need to be saved again to memory. This overhead can be described as a multiplication between the transition cost multiplied by the number of times the transition happens, as described in Equation \ref{eq:5}. Notice that larger regions tend to have entire cycles inside it, such as entire nest of loops, thus reducing the number of transitions.

\begin{equation}
\small
    Interp_{Time} = Interp_{Cost} \times InterpFreq
\label{eq:1}
\end{equation}
\begin{equation}
\small
    Native_{time} = Native_{Cost} \times  NativeFreq
    \label{eq:2}
\end{equation}
\begin{equation}
\small
    Gen_{time} = Gen_{Cost} \times HotStaticSize + CompilerInitializationCost \times NumRegions
    \label{eq:3}
\end{equation}
\begin{equation}
\small
 TransitionTime = TransitionCost \times NumTransitions
\label{eq:5}
\end{equation}
\begin{equation}
\small 
 Interp_{time} + Native_{time} + Gen_{time} + TransitionTime < Interp_{Cost} \times TotalFreq
\label{eq:4}
\end{equation}

Although these equations offer a simplified overhead model for DTEs, it gives us a significant insight into the performance of dynamic translator: the more frequently executed are the compiled regions of code, the higher will be the speedup when comparing to solely interpreting it.
Another interesting point is that this hotness characteristic is inherent to the program being emulated, not to the DTE~\cite{cad13}. 
For example, a program could execute each of its instructions only one time, being impossible to achieve any speedup with dynamic compilation.
Hence, the performance of the DTE depends also on the characteristics of the program being emulated.

The cold emulation, hot emulation, compiler, and region transition overheads define the main factors in a DTE performance. 

Hot emulation ($Native_{Cost}$) performance is directly related to the quality of the code generator by the DTE compiler. Many decisions such as the shape and size of the regions affect the quality of the code generated. 
For instance, regions with a more substantial number of instructions may expose many more optimization opportunities to the compiler, but regions with more branches are more susceptible to early exits due to phase changing~\cite{hsu2013improving}, leading to region fragmentation~\cite{Lei}, code duplication~\cite{scott2004overhead}, and infrequently executed region tails~\cite{borin2009characterization}.
Another problem with large regions comes from exception handling: given the difficulty to map exceptions during native execution, the DTE may need to roll the execution back and reinterpret the entire region every time an exception occurs and regions with frequent exceptions may become a performance issue~\cite{haubl2011trace}. Larger regions have more probability of including more branches and exceptions.  

Giving the main performance factors of DTEs and the characteristics of the compilation units chosen by an RFT to generated high-performance code, we selected seven metrics that are important when trying to better understand and analyze the performance behavior of a DTE. These metrics are described as follows:

\begin{itemize}
	\item \textbf{Total Number Of Regions}: indicates how many regions the RFT formed.
	This metric provides insights about the compilation overhead, the more frequent the compiler is called, the higher will be its overhead.
	\item \textbf{Regions Coverage}: reflects the percentage of the instructions that are being emulated by translated code, instead of interpretation ($InterpFreq/NativeFreq$).
	This metric indicates how much the hot code detection and the region formation policy are guessing correctly.
	The more instructions are executed outside the regions, the higher is the chance of existing hot code that was not included in a region.
	It is important to compare this metric with the number of regions because forming fewer regions at the cost of lower coverage is not desirable.
	\item \textbf{Number of transitions}: is the number of entrances in regions which came directly from the exit of other regions ($NumTransitions$).
	A high amount of transitions may cause a higher pressure over the processor code cache, and it is associated with the fragmentation of code cycles (nested loops, for instance).
	Furthermore, transitions over regions have an emulation cost.
	Thus, a low number of transitions imply in a good dynamic region quality.
	\item \textbf{Dynamic Region Size}: is the total number of dynamic instructions emulated by the region divided by the number of times that region was entered.
	It is important to notice that the average dynamic size of regions with low completion ratio can be smaller than its static size. 
	On the other hand, regions with loops can have a dynamic size much more prominent than their static size.
	This metric indicates the locality of execution; the more significant is the dynamic size, the lower is the number of transitions between regions.
	\item \textbf{Static Region Size}: this is the average number of instructions per region.
	Therefore, it is also correlated with the compilation overhead, as the compilation time is usually related to the number of instructions being compiled ($Gen_{Cost} \times HotStaticSize$). 
	\item \textbf{Completion Ratio}: is the percentage of times a region is executed entirely, which means that all instructions in that region were executed from the entrance to its exit.
	This metric makes more sense when dealing with superblocks, like the ones formed by NET or MRET2, which have a main exit well defined.
	Regions such as the ones formed by NETPlus do not have a clear distinction between side-exits and main-exits.
	A good completion ratio on traces means fewer early-exits, which can have a significant impact on fragmentation.
	\item{\bf 90\% Cover Set:} indicates the minimum number of regions needed to cover 90\% of the executed code frequency.
	The lower is the cover set, the fewer regions are needed to cover the hot part of the code, and these are the regions that should incur further optimizations. 
\end{itemize}

Collecting and understanding these metrics is important to understand the advantage of each RFT and DTE design choice and its drawbacks. In the following section, we show that more interesting than the metrics themselves, it is not necessary to fully implement a DTE to collect them. 
We only need to simulate the states transitions from a DTE during the emulation of a binary. To prove so, we implemented such a DTE life-cycle simulator, named RAIn, and simulated the execution of multiple applications using different RFTs.

\section{RAIn: A Dynamic Translator Simulator}
\label{sec:rain}

The implementation and evaluation of region formation techniques in a real-world DBT or HLLVM is not an easy task.
It involves debugging dynamically generated machine code among other complex tasks, which is overall a very time demanding job.
For that reason, it is seldom to see DT systems that implement more than one RFT technique on a single DBT, which is why it is very difficult to make a fair comparison among different region formation techniques.

Our approach to this problem was to develop an open-source tool, called the {\it Region Appraise Infrastructure (RAIn)}\footnote{RAIn's source code: https://github.com/vandersonmr/Rain3}, to simulate the execution of a dynamic translator, allowing an easy and flexible prototyping process.
RAIn relies on the {\it Trace Execution Automata (TEA)}~\cite{Tea} technique to mimic region formation and execution and to collect accurate region profile information. 
Initially, the TEA technique was used to record execution traces along with profile information for future executions~\cite{Tea}.
A TEA uses a Deterministic Finite Automata, or DFA, to map executed instructions or basic blocks to pre-defined traces.
RAIn also applies a DFA, but it maps instructions to regions, formed according to a pre-defined RFT.

\begin{figure*}[htb!]
\begin{subfigure}{0.485\textwidth}
	\centering
	\includegraphics[width=0.7\columnwidth]{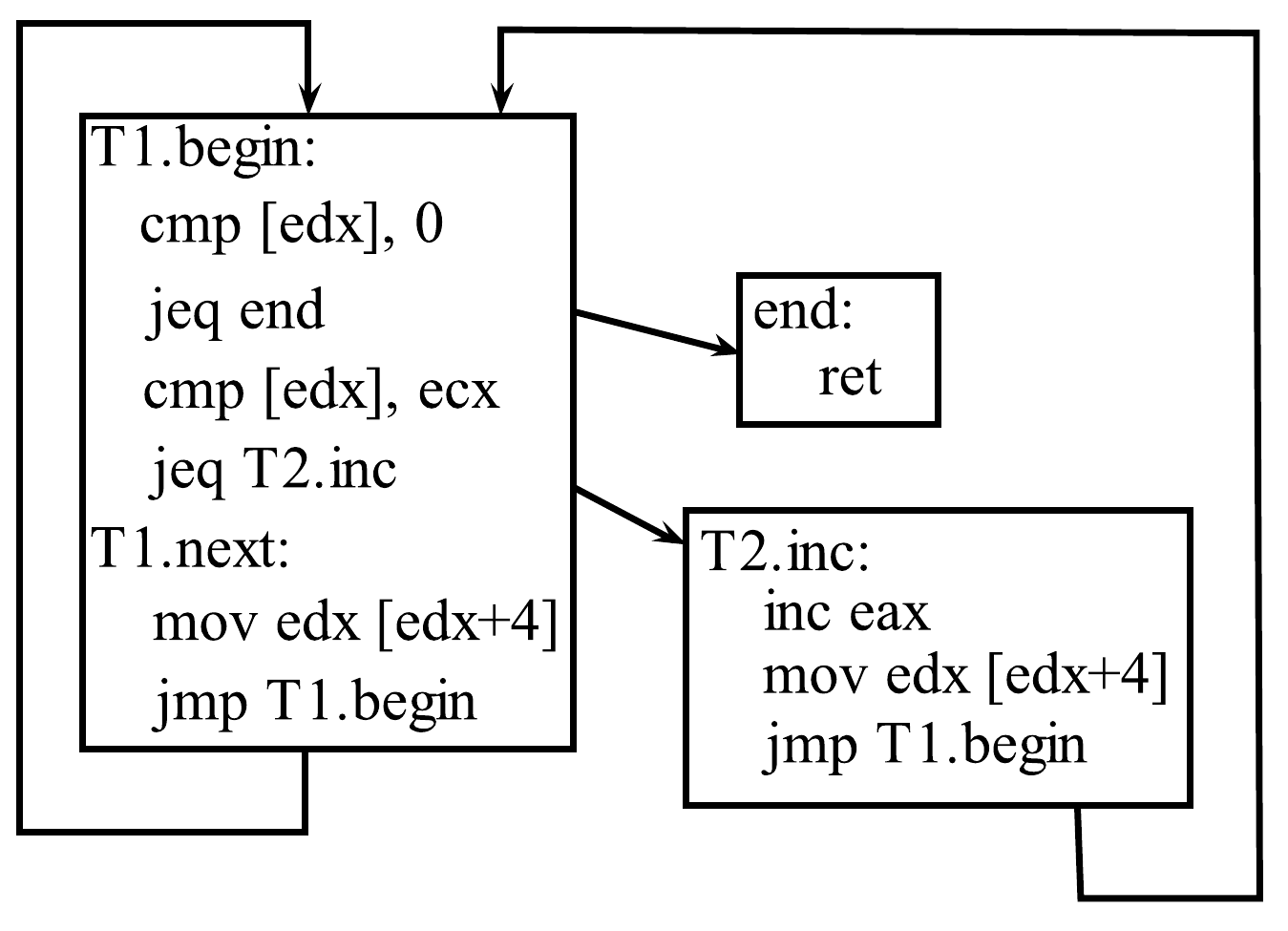}
	\caption{\label{fig:rain-dfa-a}NET superblocks}
\end{subfigure}
\hspace{0.2cm}
\begin{subfigure}{0.485\textwidth}
    \centering
	\includegraphics[width=0.7\columnwidth]{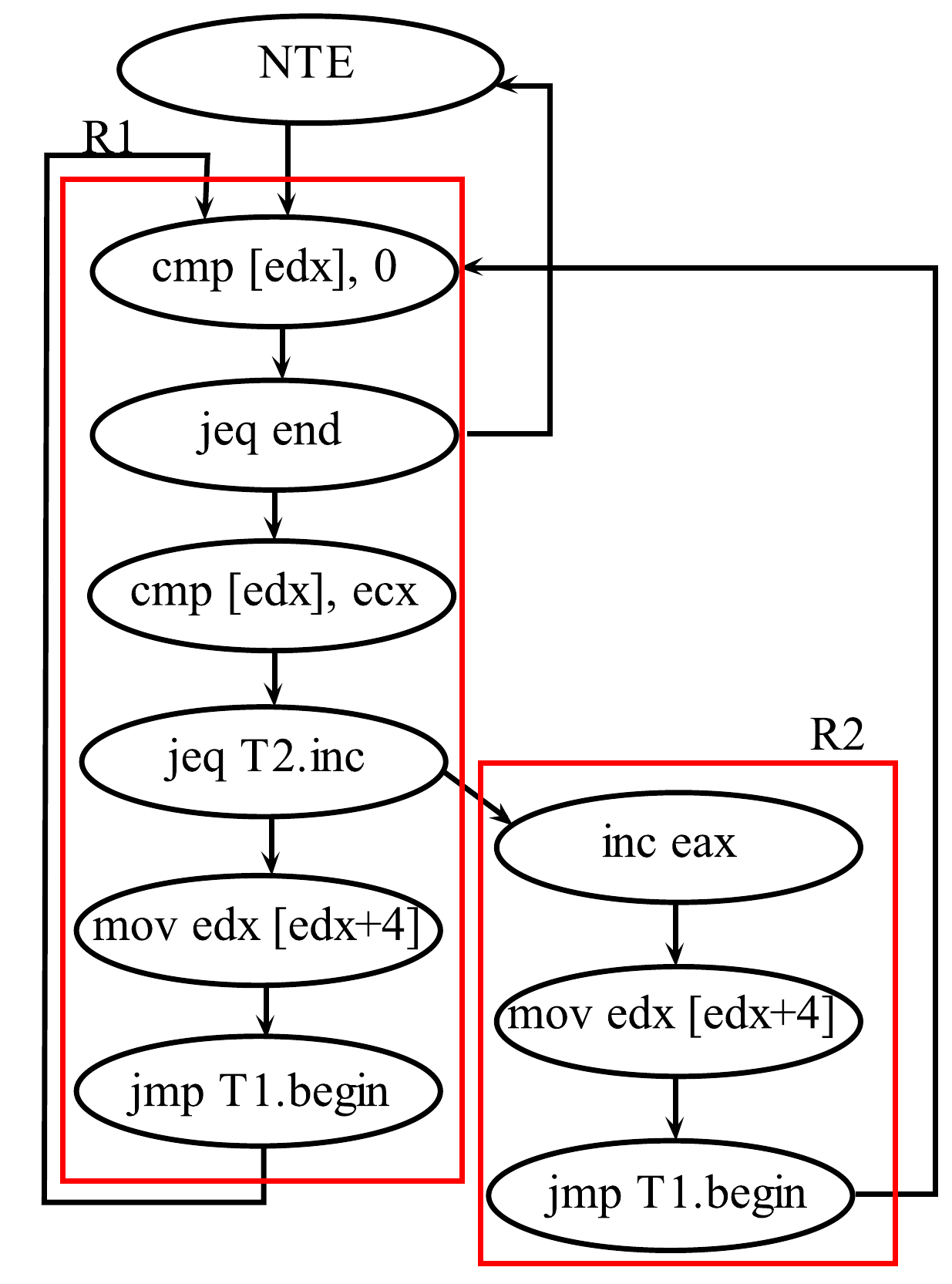}
	\caption{\label{fig:rain-dfa-b}RAIn DFA}
\end{subfigure}
\caption{Example of RAIn state blocks.}
\end{figure*}

RAIn implementation is not dependent on any architecture and it can be used to simulate a  RISC-V, ARM, or x86 input with no changes. The input to RAIn is a sequence of instructions that can be collected from the execution of a program (address and opcode), for example, using a simulator. 
It consumes this sequence of instructions, but instead of executing them, like a functional DTE, an automaton is dynamically traversed and updated, representing the execution of instructions by regions.
This automaton is also expanded under certain circumstances, representing the creation of regions.
Initially, only one state, called {\it No Trace being Executed (NTE)} is present in the automaton.
This state keeps track of instructions that belong to no regions and is used to account for instructions that are executed by an interpreter on a virtual machine that couples interpretation with dynamic binary translation, for example.
This state prevents the system from creating a new state for every single instruction executed, which could bloat the automaton.
Figure~\ref{fig:rain-dfa-b} represents a RAIn DFA that has been created when executing the trace generated by the program in Figure~\ref{fig:rain-dfa-a}.
The regions {\tt R1} and {\tt R2} were formed according to the NET RFT.
Notice that each instruction represents a state at the DFA and states are grouped into regions, representing the regions formed by the RFTs.
Edges between states crossing R1 and R2 boundaries represent transitions between the two regions.

So, whenever an instruction is consumed, RAIn checks the automaton for a valid transition leaving the current state.
If there is an outgoing edge labeled with the address of the consumed instruction, then, RAIn performs the transition, updating the current state along with the edge and state execution statistics.
If there is no outgoing edge that represents the execution of the consumed instruction, then it means that this path has not been recorded before.
In this case, a new edge is created and added to the automaton, representing a new valid transition.
This may happen due to a side-exit execution, like the transition from instruction {\tt jeq T2.inc}, on region {\tt R1}, to instruction {\tt inc eax}, on {\tt R2}, for example. 

The RFT technique monitors the automata transitions and, according to its policy, it may start the formation of a new region.
During this phase, instead of transitioning on existing states, RAIn records the executed instructions and associated transitions until it reaches the RFT stop criteria.
After reaching the stop criteria, RAIn updates the automaton, creating new states and transitions that represent the instructions and correct execution flows inside the new region.  

The operation of RAIn itself can be seen as a state machine.
Starting at the {\tt EXECUTE} state, the system consumes instructions performing transitions on the current DFA and recording statistics.
Once the RFT triggers the region formation, the {\tt RECORD} state is activated, and RAIn starts recording a new region based on the flow and the instructions being consumed.
After the RFT identifies the stop condition, RAIn enters the {\tt APPEND} state, in which it expands the DFA with the newly formed region.
Once the DFA is expanded, the system returns to the {\tt EXECUTE} state, continuing with the automata execution.

RAIn is implemented in two modules: the {\it RegionManager} and the {\it Simulator}.
The {\it RegionManager} is the module responsible for managing the policies for RFTs.
It controls the start and stops criteria for region recording.
To add a new RTF to RAIn, all that is necessary to provide is an implementation of a {\it RegionManager} and a respective call in the main function to register it on the \textit{Simulator} module. 
Every \textit{RegionManager} implements the method ``handleNewInstruction'' that is called for every instruction from the trace and it should handle region creations. Code \ref{lst:NET} shows the whole implementation of the NET RFT. With less than 20 lines of code we implement and RFT and are able to analyze it using RAIn's metrics with instructions traces from different ISAs and Operational Systems.  

\renewcommand{\lstlistingname}{Code}
\lstset {language=C++,basicstyle=\footnotesize}
\begin{lstlisting}[basicstyle=\tiny, label=lst:NET,caption=NET Implementation using RAIn]
Maybe<Region> NET::
handleNewInstruction(trace_item_t &LastInst, trace_item_t &CurrentInst, State LastTransition) {
  if (Recording) {
    if (wasBackwardBranch(LastInst, CurrentInst) || LastTransition == InterToNative) {
      Recording = false;
      return Maybe<Region>(RecordingRegion);
    }
    RecordingRegion->addAddress(CurrentInst.addr);
  } else if ((LastTransition == StayedInter && wasBackwardBranch(LastInst, CurrentInst)) ||
             LastTransition == NativeToInter) {
      HotnessCounter[CurrentInst.addr] += 1;
      if (isHot(HotnessCounter[CurrentInst.addr])) {
        Recording = true;
        RecordingRegion->addAddress(CurrentInst.addr);
        HotnessCounter[CurrentInst.addr] = 0;
      }
  }
  return Maybe<Region>::Nothing(); 
}
\end{lstlisting}

RAIn processes a sequence of instructions that represent the execution of a program.
RAIn processes one instruction at a time, similar to an interpreter. 
Hence, its performance is similar to the performance of an interpreter when evaluating a single RFT.
In case the user aims at evaluating several RFTs, she may parallelize the simulation by loading the sequence of instructions once and feeding several RAIn threads, each one simulating a different RFT or hyper-parameter.
We employed this approach to collect the results from our experiments and it took near to half an hour to collect all statistics from all tested RFTs from each benchmark trace with 10 billion-instruction in an Intel Xeon E5-2630 (2.60GHz).

\section{Experimental Setup}
\label{sec:setup}

To evaluate RAIn, the study presented in this paper was conducted using applications from two benchmark suites (\SPECCPUsix\ and \SYSMARKtwe) and a Linux-compatible and open-source image editor, GIMP.
A total of 14 benchmarks from \SPECCPU~\cite{SPEC2006} were applied in this study, ten from SPEC-FP and four from SPEC-INT; and four benchmarks from \SYSMARK~\cite{Sysmark}.

The usual benchmark suite for RFT related research in the literature is \SPECCPU.
However, \SPECCPU\ and \SYSMARK\ have a noticeably different profile, and one of the goals is to understand how these DTEs configurations perform across all these types of applications.
\SYSMARK\ is described as {\it an application-based benchmark that reflects usage patterns of business users in the areas of office productivity, data/financial analysis, system management, media creation, 3D modeling, and web development}.
In this work, we have evaluated the effect of RFT techniques on office applications, combining four applications from the \SYSMARK\ Office scenario (FineReader Pro 10.0, Internet Explorer 8, PowerPoint 2010, and Word 2010) and GIMP (2.8.20).
This set of applications form what we will call {\it Desktop Apps}, and aims to represent a group of applications with a large 90\% Cover set, as opposed to the \SPECCPU\ benchmarks.

Several applications from both of these benchmark suites generate sequences with trillions of executed instructions.
The chosen method to handle such amount of data was to use RAIn to simulate 10 billion-instruction sequences of each program.
We executed all these benchmarks on Bochs Emulator~\cite{lawton1996bochs} and captured the x86 executed instructions to form the sequence.
To avoid initialization code, the first 10 billion instructions were discarded, and then the next 10 billion were recorded.
This number of instructions proved to be enough to expose the differences in the behavior of applications and RFTs.
These can be seen in the huge variation of code execution locality demonstrated by the 90\% cover set, and the number of basic blocks presented in Section~\ref{sec:overheads_new} and the RFT behavior difference showed in Section~\ref{sec:results}.
\section{Experimental Results}
\label{sec:results}

\subsection{\textbf{Parameter Selection}} 

There are two important parameters in our tested RFTs: hotness threshold and NETPlus deepness.
To selected a good value, we ran two benchmarks and tested several parameter values.

\begin{figure*}[h!]
    \centering
    \includegraphics[width=0.9\textwidth]{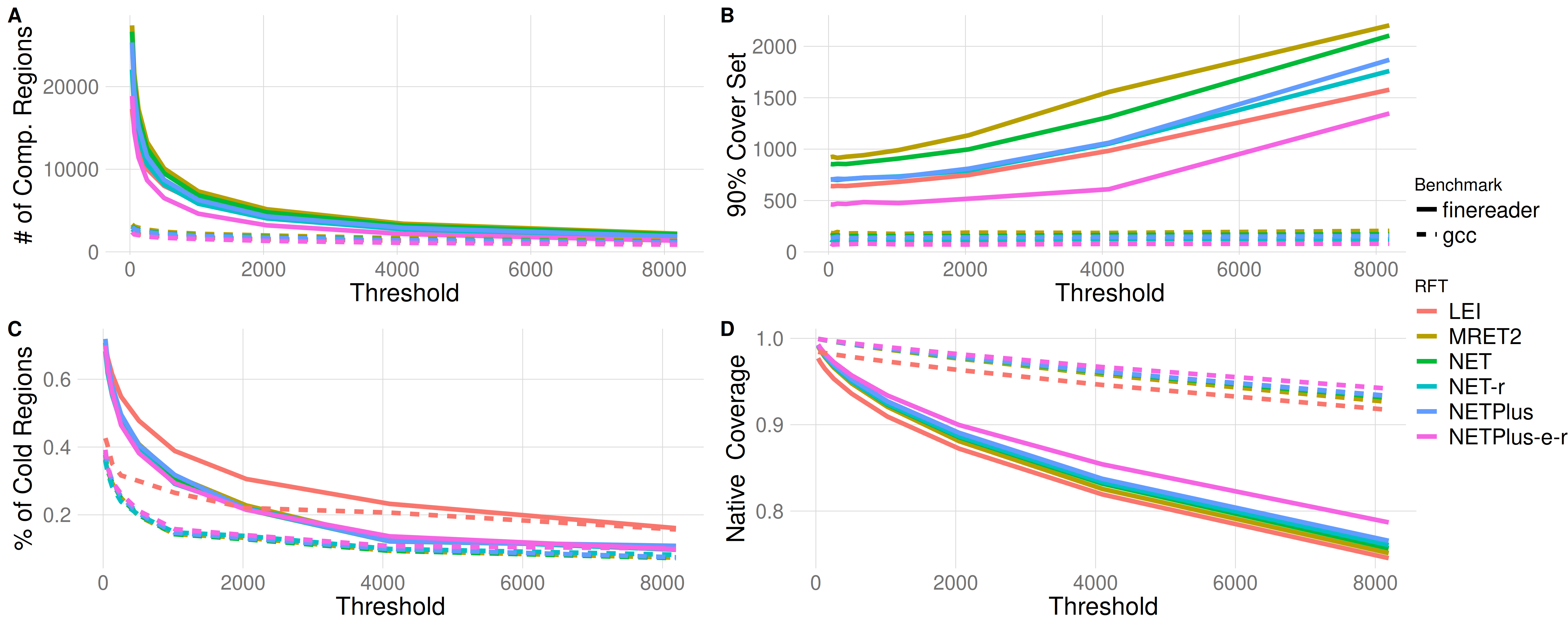}
    \caption{Impact of region hotness threshold on the A) 90\% Cover Set, B) number of compiled regions, C) native execution coverage, and D) percentage of cold regions for six RFTs.
        The data was generated using 10-billion-instruction sequences from Finereader (\SYSMARK) and GCC (\SPECCPU) benchmarks.}
    \label{fig:threshold}
\end{figure*}

\textbf{Threshold Value}\\
Figure~\ref{fig:threshold} shows how the number of compiled regions, the 90\% cover set, the percentage of cold regions, and hot-emulation coverage are affected by the hotness threshold. 
As we can notice from these graphics, a slight variation of the threshold value can significantly decrease the number of basic blocks selected for translation, reducing the compilation overhead.
The same effect occurred when varying the threshold of all the RFTs, as we can observe in Figure~\ref{fig:threshold}(A).
We can also observe that by choosing a threshold near 1024, we got a low number of regions selected and cold region proportion without losing too much hot-emulation coverage (only when the threshold is near 2000 the hot-emulation coverage becomes less than 90\% for Finereader). Thus, we set 1024 as a fixed threshold for all the next experiments.

As depicted in  Figure~\ref{fig:threshold}, all the tested RFTs are strongly influenced by the threshold and increasing its value not only reduced the proportion of cold regions, proving the strong correlation between the past execution frequency and its future, but also increased the 90\% Cover Set value.
Therefore, the threshold value has a large influence over the four metrics and so, its choice should not be neglected or ignored in the design and construction of a DTE.

\begin{figure}[h!]
 \centering
    \includegraphics[width=0.7\columnwidth]{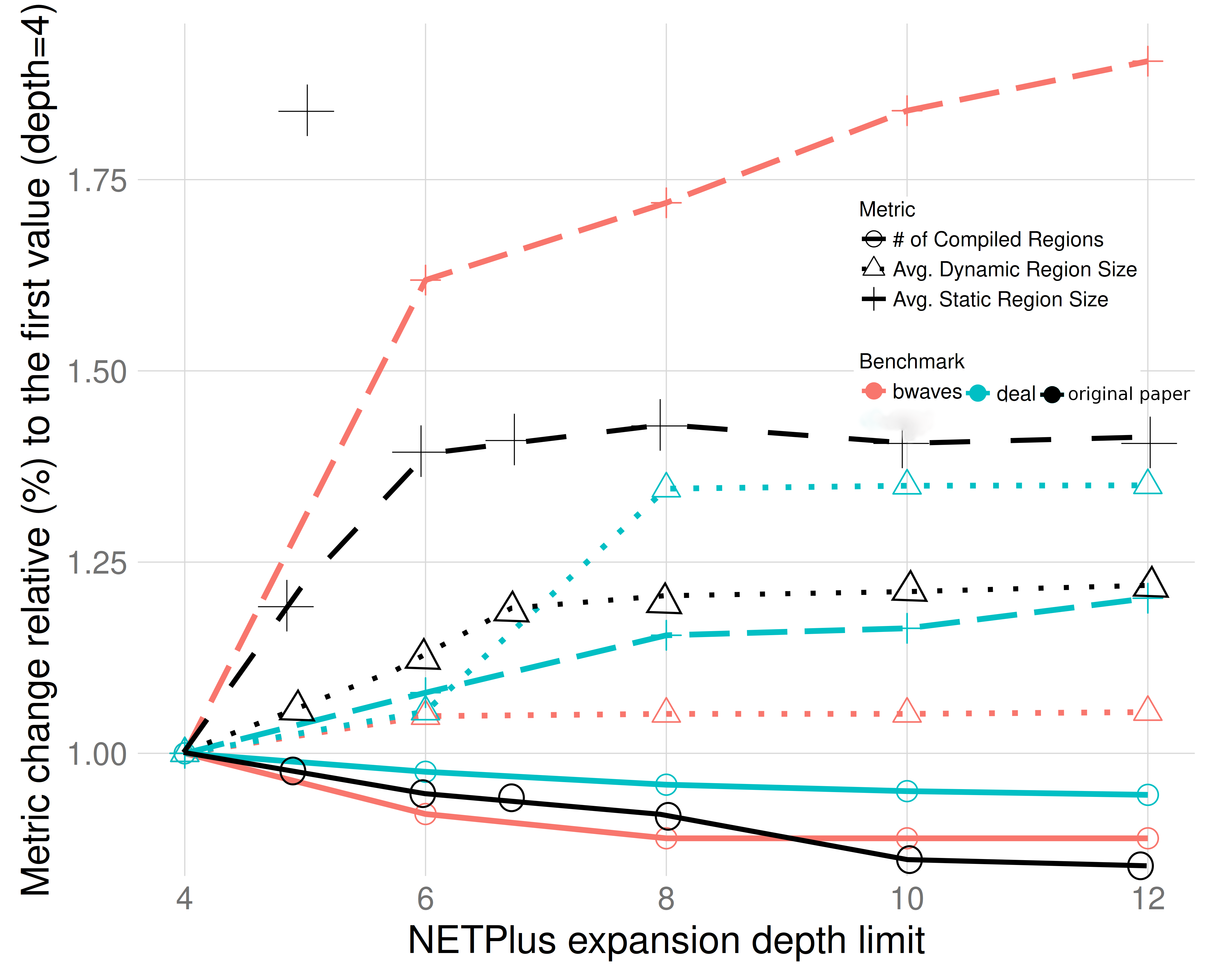}
    \caption{Impact of the NETPlus expansion depth limit on the number of compiled regions, avg. dynamic region size and avg. static size variation for two \SPECCPU\ applications (we choose from SPEC to be easier to compare with results from the NETPlus original paper). The results are normalized by the first value (depth = 4).}
    \label{fig:netplusdepth}    
\end{figure}

\textbf{NETPlus Deepness}\\
As explained in Section~\ref{sec:trace-based}, the NETPlus RFT has an expansion depth limit that controls how far the search for loops in the original NET region can go.
Figure~\ref{fig:netplusdepth} shows how the number of compiled regions, the average dynamic region size, and the average static region size are affected by the NETPlus expansion depth limit.
The graphic shows that there is a stabilization in the metrics when the depth limit gets near to ten and, thus, choosing a value higher than ten would probably have no benefit.
Hence, we fix the NETPlus expansion depth limit in ten for all the following experiments.

One important observation is that the results in Figure~\ref{fig:netplusdepth} are very close to the ones presented by the authors of NETPlus~\cite{Netplus}, demonstrating the capabilities of RAIn to explore RFT properties and leading its users to obtain the same conclusions as for when using a real DTE.
Additionally, Figure~\ref{fig:netplusdepth} shows that the increase in the average static size of the regions made by the NETPlus expansion can be much more costly for some programs than for others, such as the case of {\it bwaves} and {\it deal}, chosen for being in different parts of the spectrum of the 90\% cover set metric. 
A high increase in the static size from bwave did not lead to any significant increase in the dynamic region size, showing that NETPlus, for some programs, can add costs that may never be paid-off, a fact, and information that was not first observed by its authors.

Hence, all the following results were generated with a hotness threshold set to 1024, a NETPlus expansion depth limit set to ten and with all benchmarks bars ordered by the 90\% Cover Set values presented in next subsection ( Section~\ref{sec:overheads_new}).

\subsection{Application Impact on DT Overhead}
\label{sec:overheads_new} 

\begin{figure*}[h!]
 \centering
\begin{subfigure}{0.7\textwidth}
	\centering
	\includegraphics[width=\columnwidth]{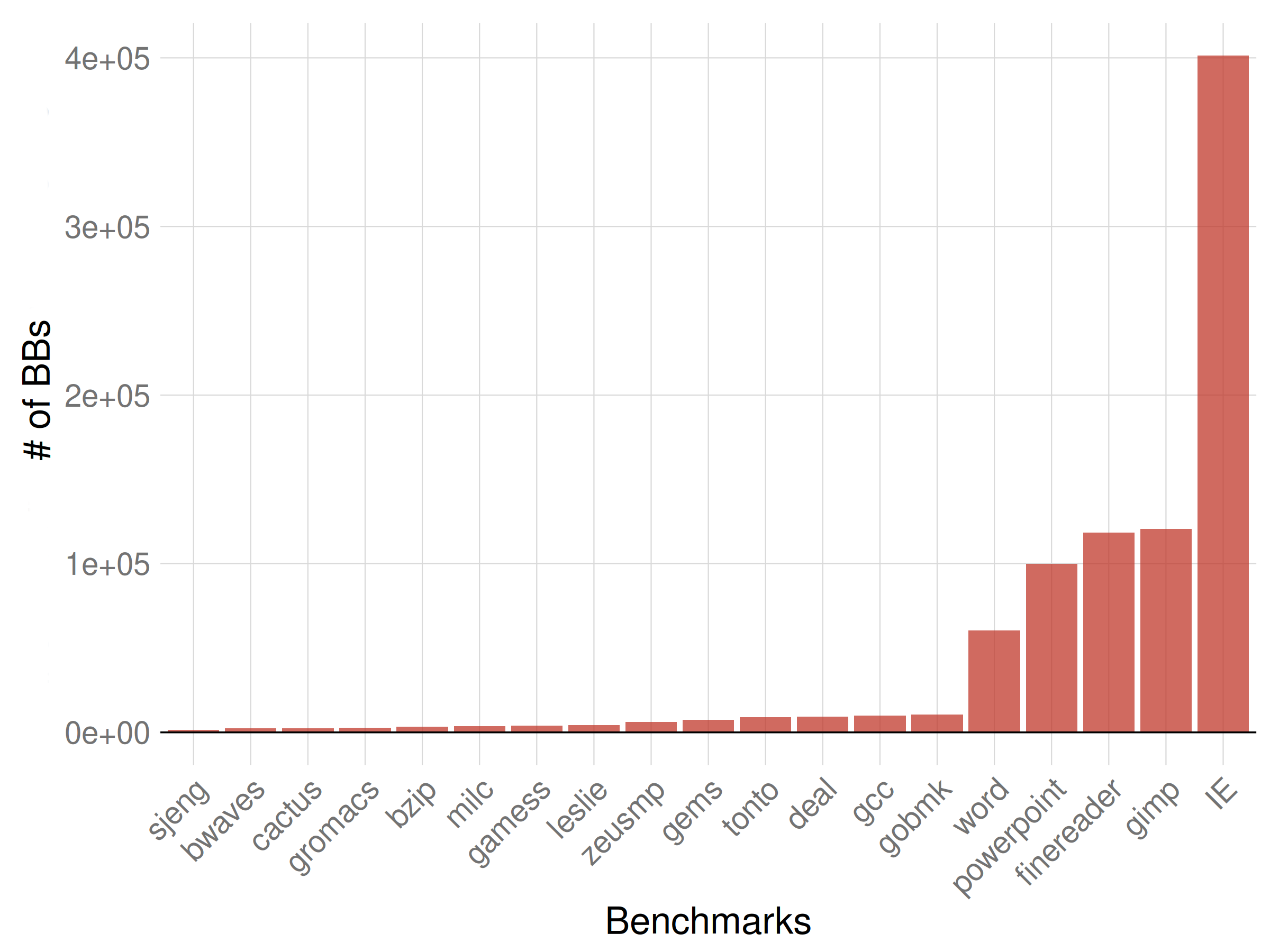}
	\caption{\label{hotbbsFig}}
\end{subfigure}
\hspace{0.2cm}
\begin{subfigure}{0.7\textwidth}
    \centering
	\includegraphics[width=\columnwidth]{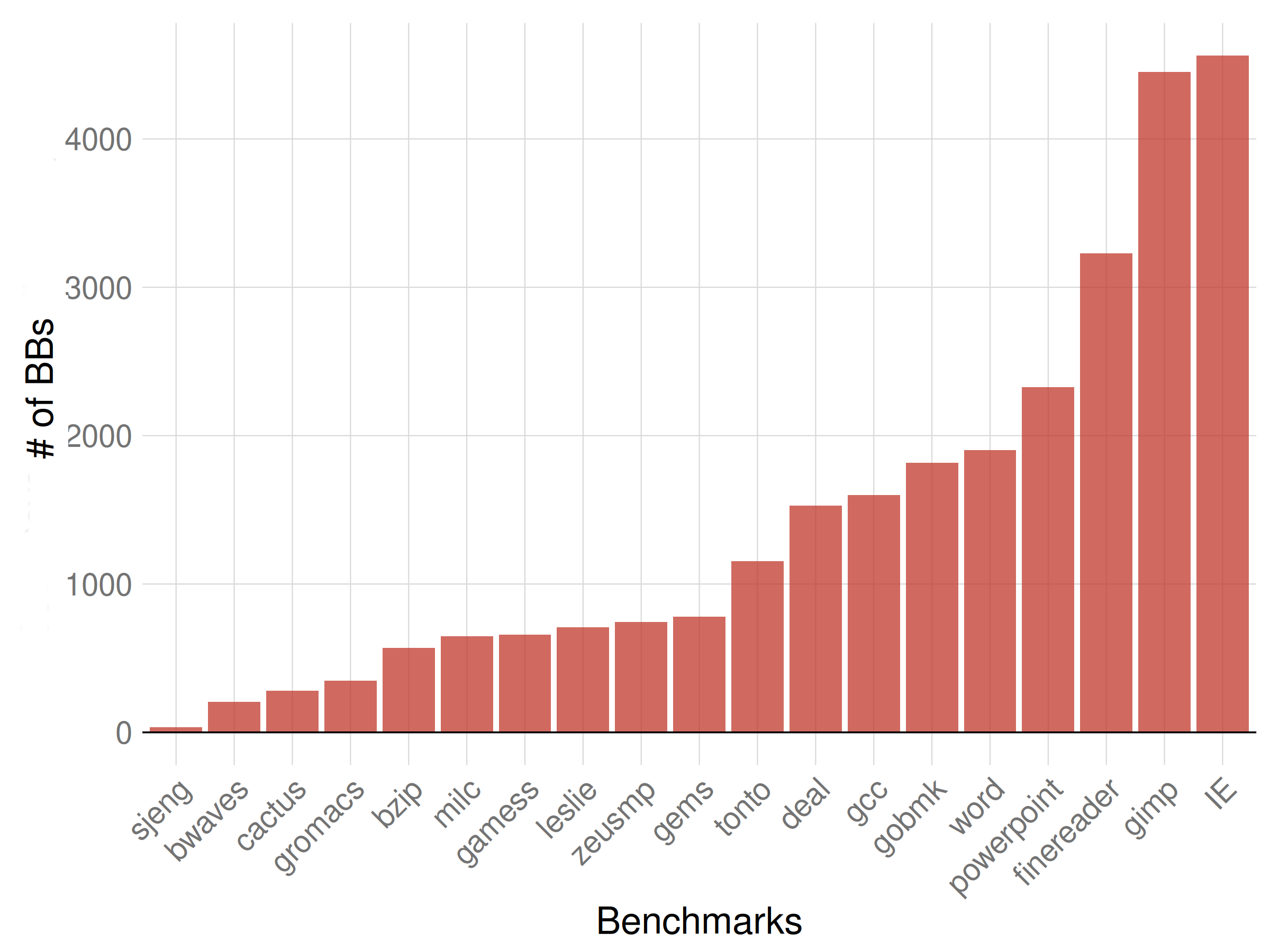}
	\caption{\label{90coversetFig}}
\end{subfigure}
\caption{(a) Number of basic blocks that execute 1024 or more times and (b) minimum number of basic blocks required to cover 90\% of the 10 billion instructions simulated per application. }
\end{figure*}

To evaluate the impact that different RFTs have on the metrics discussed in Section~\ref{sec:overheads}, we execute RAIn with sequences of x86 instructions extracted from the selected benchmarks. 
For each application, we skip 10 billion instructions and then record the next 10 billion ones.
We also added a sequence of instructions from a Linux Image Editor, GIMP, to compose our set of desktop applications. 
In our tests, we considered that only basic blocks executed more than 1024 times are hot enough to be worth compiling.
As can be seen in Figure~\ref{hotbbsFig}, some applications have very few basic blocks that reach this frequency, while others have a lot of them. 
As the total execution frequency for each presented program is constant (10 billion x86 instructions), having more basic blocks with high execution frequency means that their average execution frequency is smaller. 
Notice that desktop applications had the less hot basic blocks which are similar to the conclusion in the work undertaken by Cesar et al.~\cite{cad13}, where they argue the importance of having office/GUI benchmarks when evaluating a DTE and argue that the low execution frequency average of these applications is a barrier to DTE's performance. 
This is one of the main reasons we included desktop applications from \SYSMARK and GIMP in our experimental setup for evaluating RAIn.

Another straightforward way to verify this is by using the 90\% Cover Set metric, first introduced by the authors of Dynamo~\cite{Bala1999}.
The 90\% Cover Set counts the minimum number of regions (in this example, basic blocks) necessary to achieve 90\% of the execution frequency.
The smaller the 90\% Cover Set, the lesser the amount of code to be compiled, and the higher the average execution frequency of these basic blocks.
Duesterwald et al.~\cite{Bala1999} demonstrated that there is a strong inverse relationship between the 90\% Cover Set size and a DTE's performance.
Therefore, it would be challenging to obtain the same performance on benchmarks with far different numbers in the 90\% Cover Set.
Some examples are the \textit{sjeng} and \textit{IE}, as we can see in Figure~\ref{90coversetFig}. 

\subsubsection{\textbf{Completion Ratio}}

\begin{figure*}[h!]
 \centering
\begin{subfigure}{0.7\textwidth}
    \centering
    \includegraphics[width=\columnwidth]{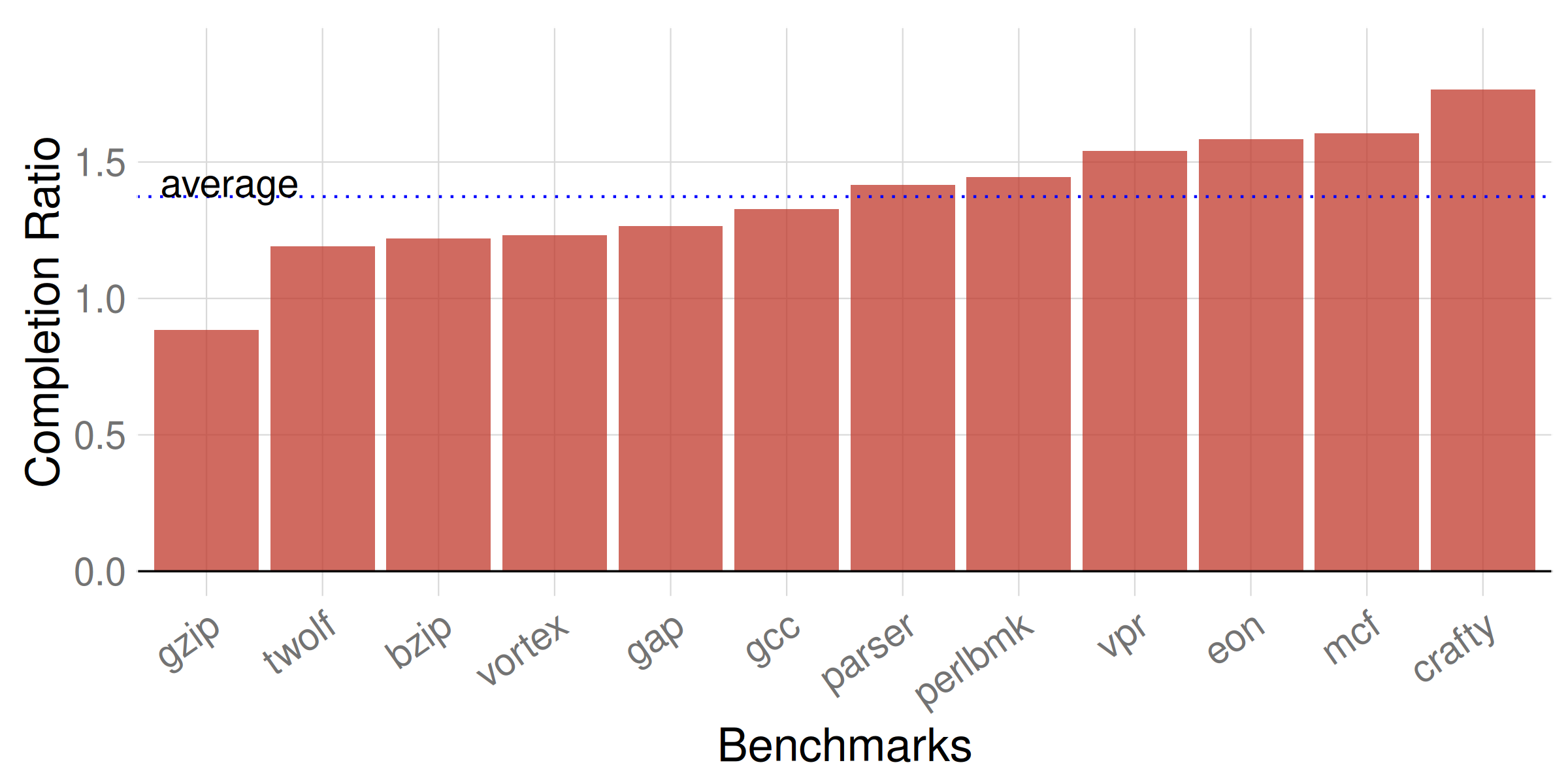}
    \caption{Data from the MRET2 patent~\cite{Mret2} -- Full execution of SPEC98 applications.}
    \label{fig:mret2-original}
\end{subfigure}
\hspace{0.2cm}
\begin{subfigure}{0.7\textwidth}
    \includegraphics[width=\columnwidth]{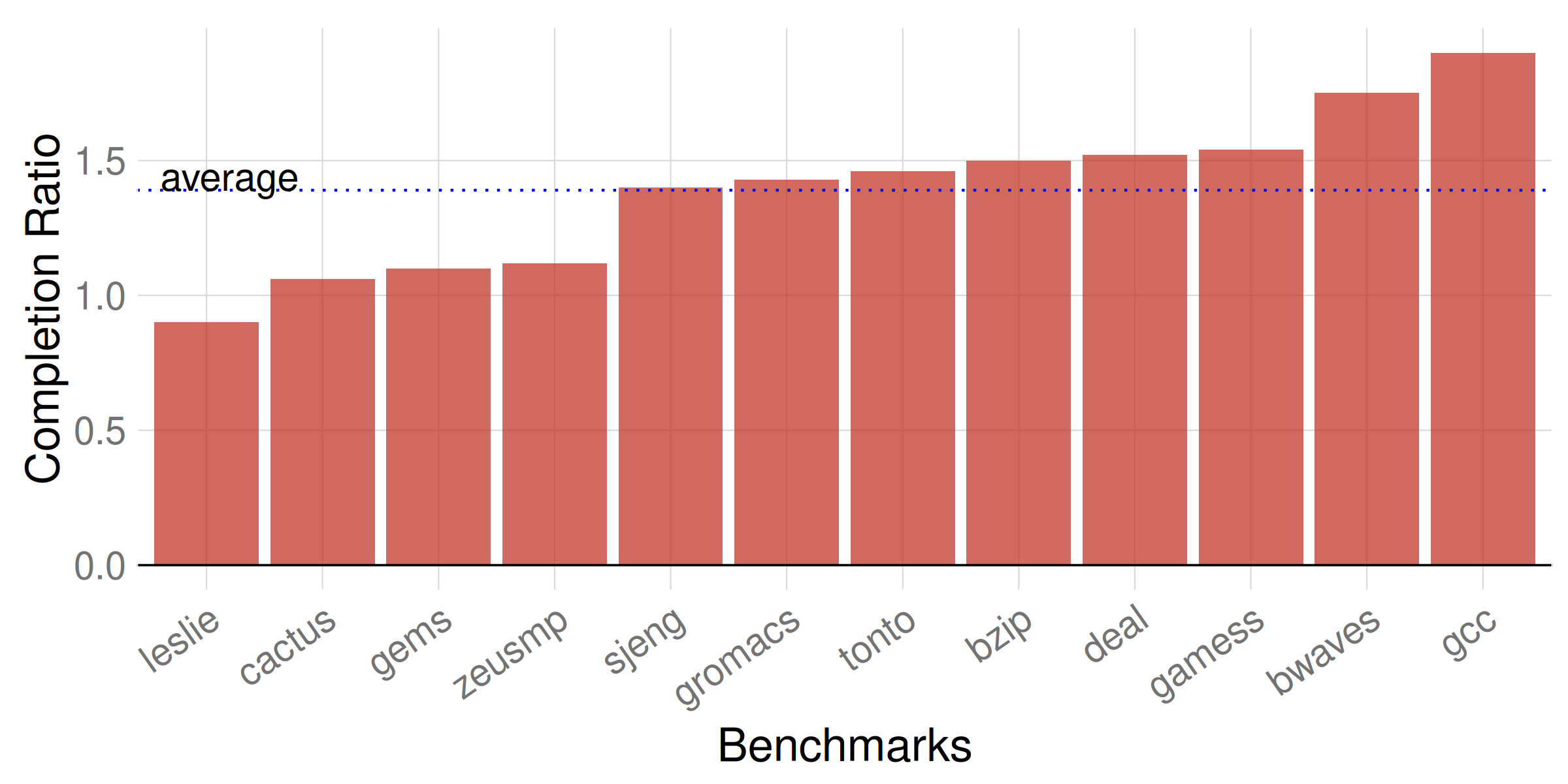}
    \centering
    \caption{Data generated by RAIn -- 10 billion instructions from \SPECCPU\ applications.}
    \label{fig:mret2-ours}   
\end{subfigure}
\label{fig:RAInDFA}
\end{figure*}

The authors of MRET2~\cite{Mret2} argue that its main advantage is the increase in the completion ratio of the selected traces over NET. 
They measured the completion ratio with MRET2 and NET with the full execution of applications from SPEC CPU 98 and show that, on average, MRET2 improves the completion ratio by 20\%. 
Figure~\ref{fig:mret2-original} shows a re-plot of their data, while
Figure~\ref{fig:mret2-ours} shows the data collected with RAIn for \SPECCPU\ 2006 applications. 
Besides the difference in methodology and benchmarks, both RAIn and the original paper present very similar results (distribution and average), leading to the same conclusions, showing again that the simulation performed by RAIn is capable of producing results similar to the ones obtained with real DTEs.

\begin{figure}[htb!]
    \centering
    \includegraphics[width=0.7\columnwidth]{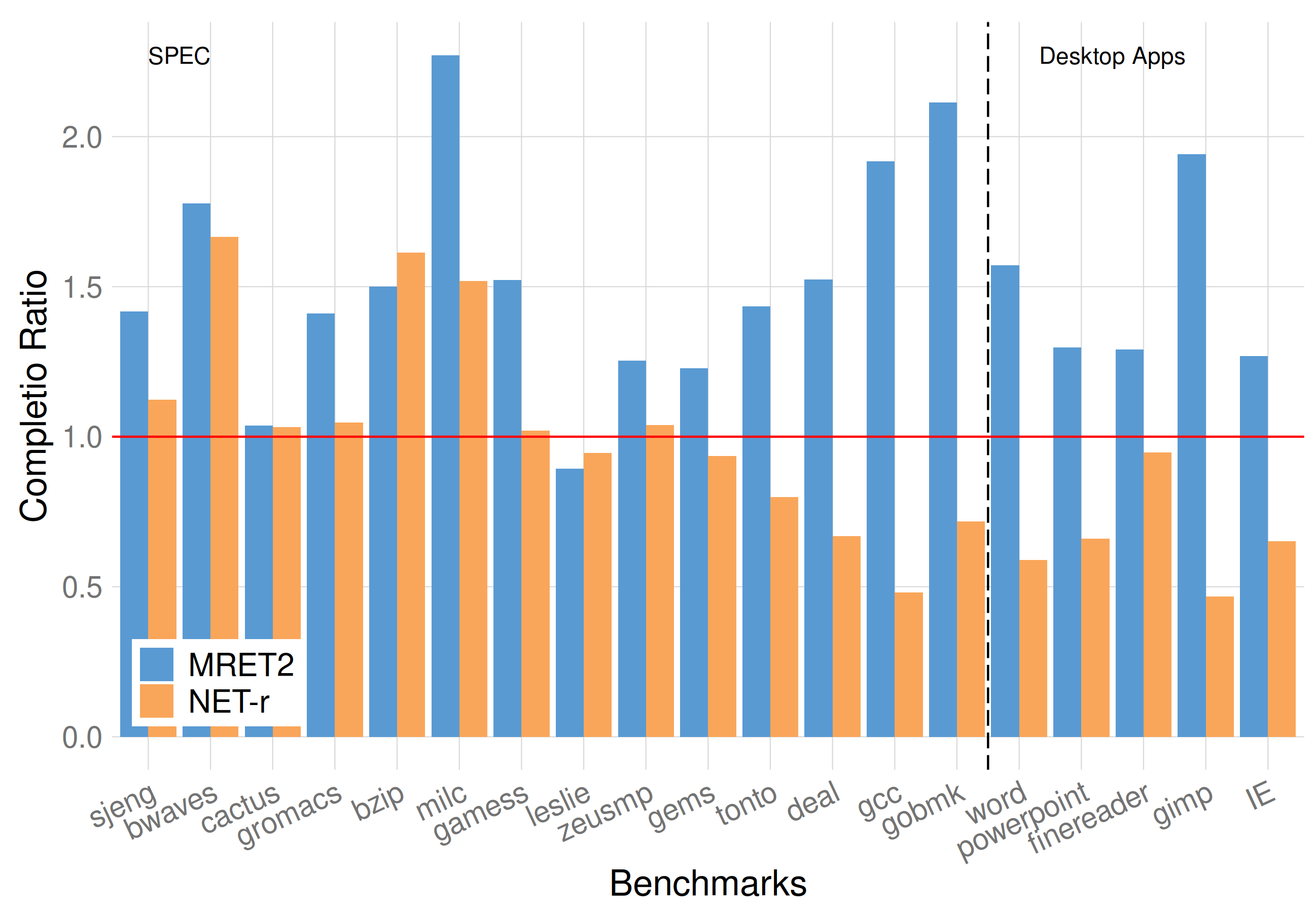}
    \caption{MRET2 and NET-r normalized completion ratios.}
    \label{fig:completion}
\end{figure}

We extrapolate the experiment using RAIn to compare NET, MRET2, and NET-R.
The results in Figure~\ref{fig:completion} show that NET-R is only better than NET in benchmarks with low 90\% Cover Set (highly dense execution frequency), i.e., the ones more to the left, while MRET2 is better than NET in almost every benchmark.

\subsubsection{\textbf{Compilation Overhead}}

\begin{figure*}[htb!]
 \centering
\begin{subfigure}{0.7\textwidth}
    \centering
    \includegraphics[width=\columnwidth]{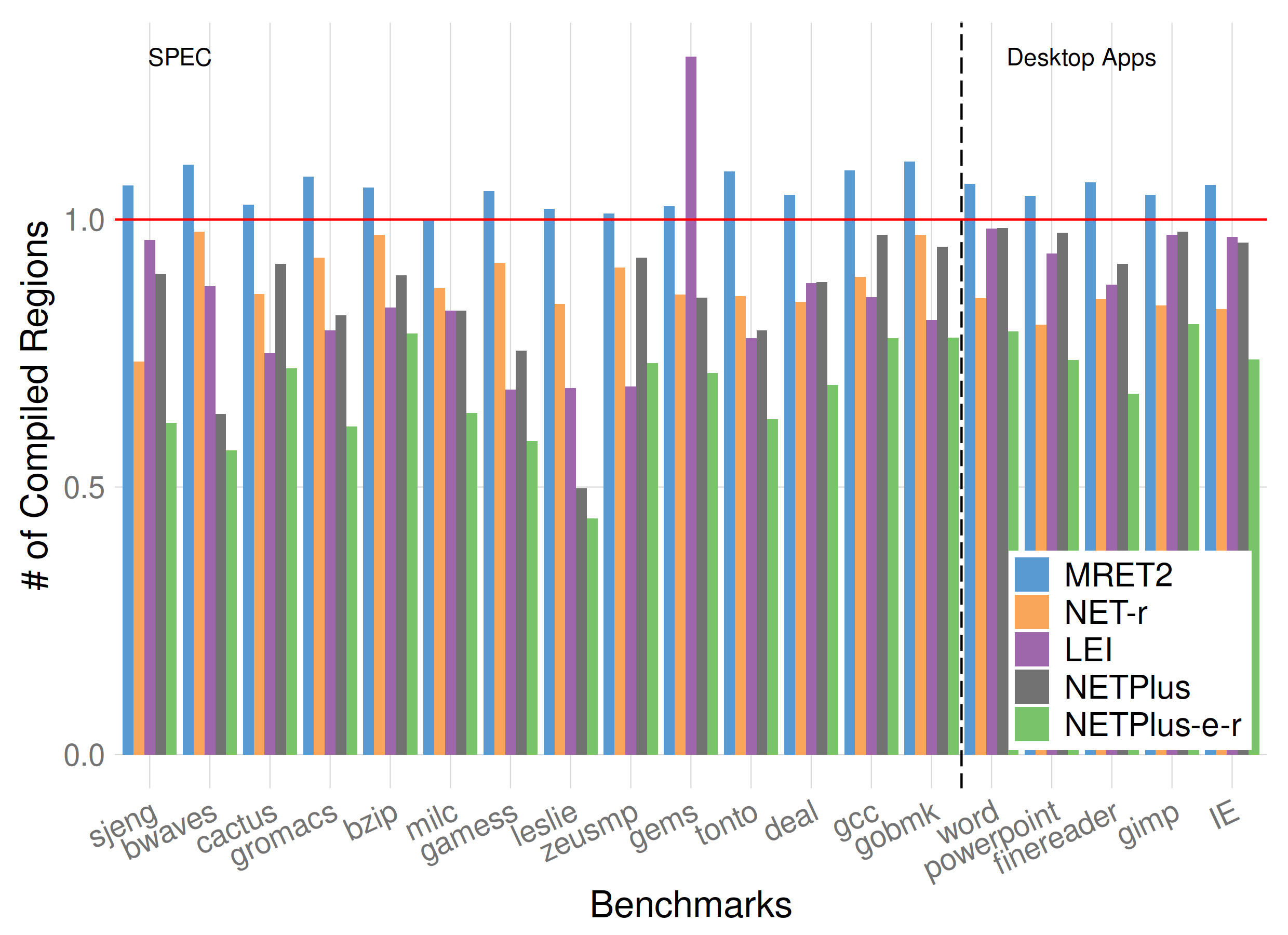}
    \caption{}
    \label{fig:numofregions}
\end{subfigure}
\hspace{0.2cm}
\begin{subfigure}{0.7\textwidth}
    \includegraphics[width=\columnwidth]{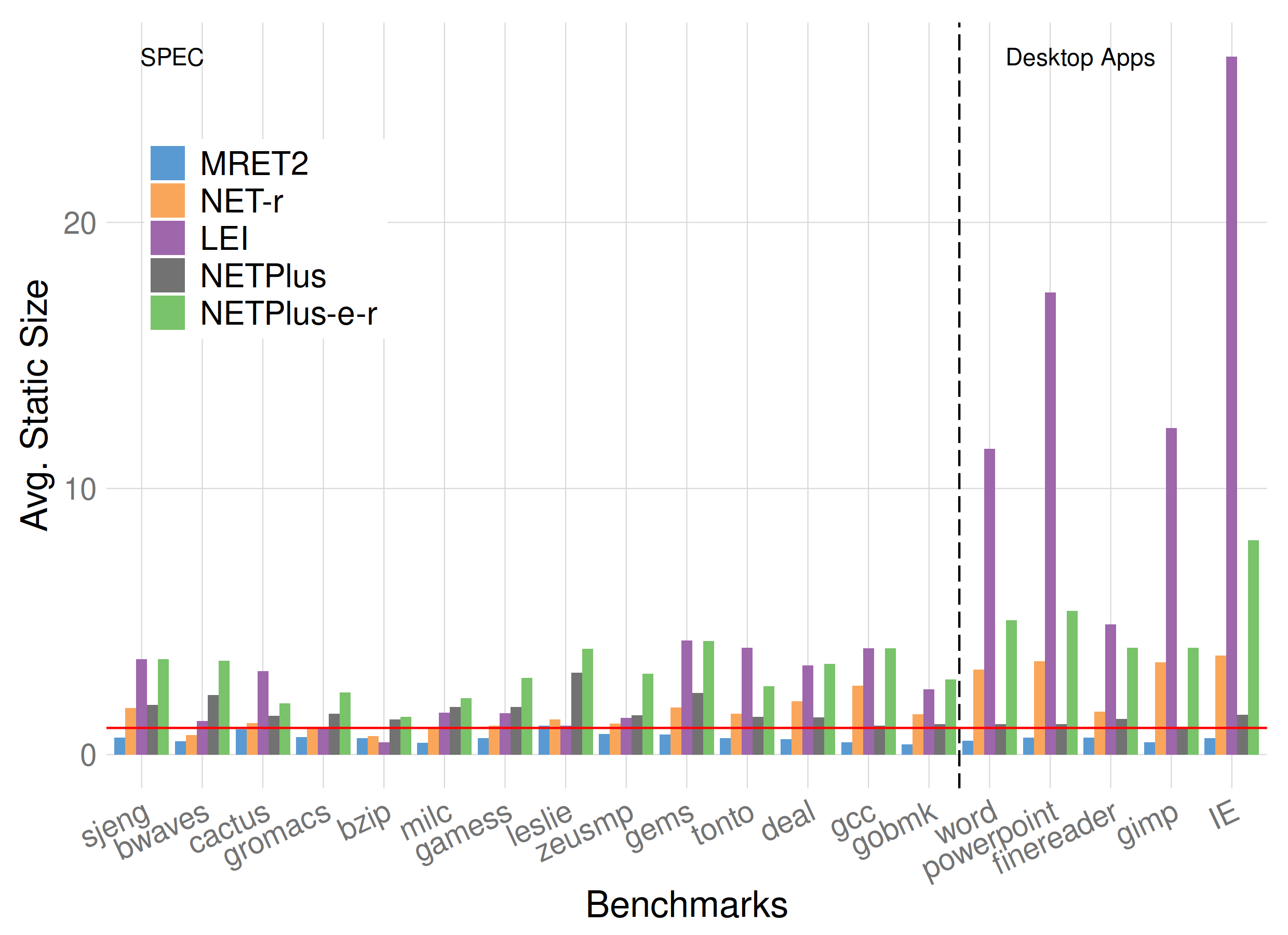}
    \centering
    \caption{}
    \label{fig:staticsize}    
\end{subfigure}
\caption{(a) Total number of Regions Compiled normalized by the NET values; and (b) the average static region sizes normalized by the NET results for of all RFTs and benchmarks.}
\end{figure*}

Since the compilation cost is correlated with the number of times the dynamic compiler is invoked and also with the number of instructions present in the compiled regions, we use RAIn to evaluate the number of compiled regions (Figure~\ref{fig:numofregions}) and the average static region sizes (Figure~\ref{fig:staticsize}).
NETPlus-e-r produced smaller amounts of regions to be compiled. 
However, despite being a much more simplistic RFT than NETPlus-e-r, NET-r had a very significant impact on this metric too. 
MRET2 and NET produced many more regions to be compiled, a result that is explored and explained by the authors of the LEI technique~\cite{Lei}.

We can also observe that there is a trade-off between the number of regions compiled and the average static region size: the majority of RFTs that decrease the number of compiled regions, also increase the average static region size.
NETPlus have the best trade-off for these metrics; it decreased the number of regions and only slightly increased the average static size;
LEI, on the other hand, has the worst trade-off.
Furthermore, notice that NET-r, NETPlus-e-r, and LEI create larger regions when emulating applications with larger 90\% Cover Sets, such as the desktop applications. 
Pointing that the relaxation of NET (NET-r) and also the expansion of NETPlus (NETPlus-e-r) have a different impact on benchmarks with different 90\% Cover Set.

\newpage

\subsubsection{\textbf{Dynamic Characteristics}}

\begin{figure*}[h!]
 \centering
\begin{subfigure}{0.7\textwidth}
    \centering
    \includegraphics[width=\columnwidth]{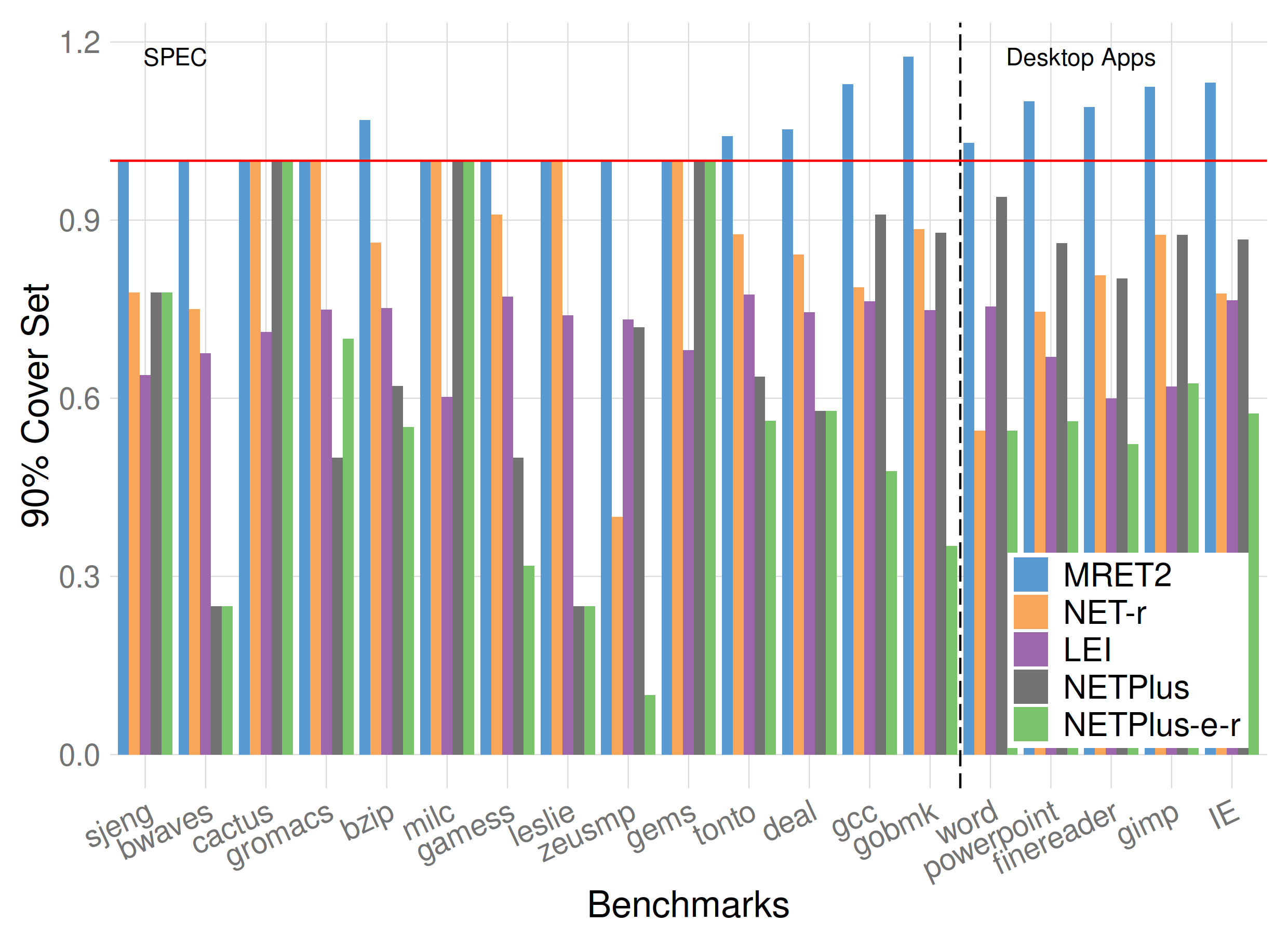}
    \caption{}
    \label{fig:90coverset}
\end{subfigure}
\hspace{0.2cm}
\begin{subfigure}{0.7\textwidth}
    \centering
    \includegraphics[width=\columnwidth]{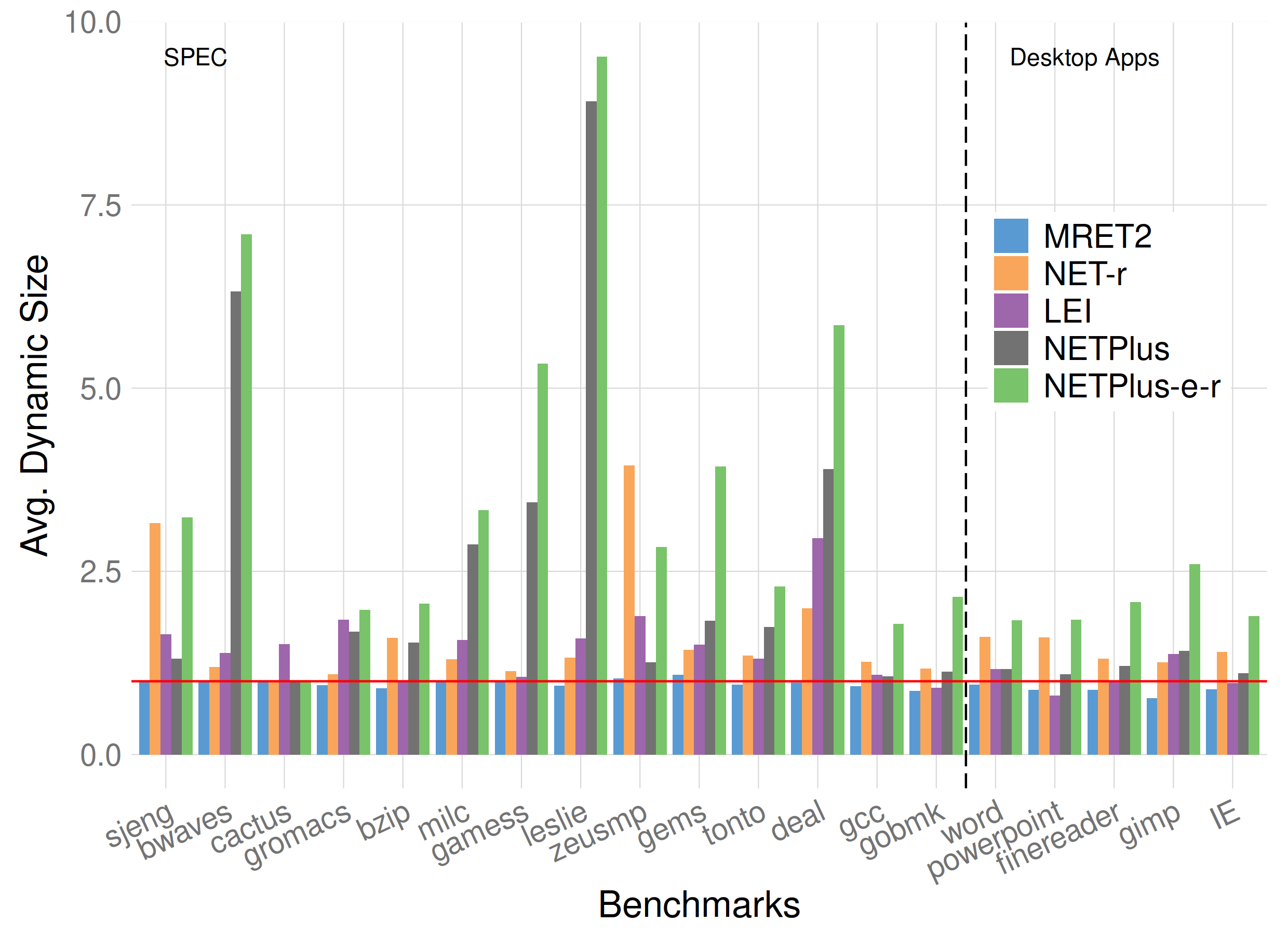}
    \caption{}
    \label{fig:dynsize}
\end{subfigure}
\caption{(a) the 90\% Cover Sets normalized by the NET results; and (b) the average dynamic region sizes normalized by the NET results for all RFTs and benchmarks.}
\end{figure*}

We also investigated the dynamic characteristics of the regions formed by all RFTs with RAIn.
Figure~\ref{fig:90coverset} shows the 90\% cover set for all RFTs normalized by the results of NET. 
In this metric, only MRET2 had a worse performance than NET, indicating that it requires more regions to be compiled to cover 90\% of the execution.
NETPlus-e-r achieved the best results, followed by NETPlus, LEI, and NET-r, with NETPlus being more efficient on benchmarks with a higher 90\% Cover Set. 
A similar result was obtained with the average dynamic size, shown in Figure~\ref{fig:dynsize}, with NETPlus-e-r achieving again the best results, followed by NETPlus and LEI, while NET-r had only a slight improvement and MRET2 decreased when compared to NET.
Last, we can notice that the results were much less significant in benchmarks with a high 90\% Cover Set.
Overall, these cases support the view that it is more difficult to select regions better than NET when the 90\% Cover Set is high.

These results are similar to the ones found by previous work, which supports our claim that using simulation to evaluate RFTs for a dynamic translator is a sound approach. 
Also, these results show that the relaxation and extension proposed by Hong et al.~\cite{hong2012hqemu} is a simple and powerful technique that should be considered when designing and implementing a dynamic translator.

\section{Conclusions}
\label{sec:conclusion}

In this work, we presented a novel DTE simulator called RAIn, which is capable of reproducing several results in the literature through simulation. The simulation enables the test of multiples DTEs’ designs, producing several useful statistics without complex implementations. Therefore, opening the new opportunities for exploration of DTE design decisions in a faster and simple manner. 

As far as we know, there is no other simulation framework for testing DTE designs like the one proposed in this paper, hence no other work similar to this one.
Moreover, there is no additional comparative study involving several RFTs whatsoever.
Typically an article that presents a new RFT only compares it with only one other more~\cite{Mret2, Lei, Netplus}, usually NET.
Finally, but not less important, the results that we obtained with RAIn in this work corroborate the findings reported by other authors in previous works. 

For example, we found with RAIn that MRET2 has a better completion rate than NET, the same result presented by the MRET2's original paper~\cite{Mret2}.
We also showed that NET and MRET2 compile far more regions than the other techniques and that these regions are smaller, a phenomenon that the LEI authors called region fragmentation~\cite{Lei}.
Furthermore, our results with the NETPlus depth limit variation showed the same graphic pattern as in the one in the NETPlus original paper~\cite{Netplus}.
Moreover, the point of convergence found by us with RAIn for the depth limit is the same as the one found by its authors.
Finally, we concluded that NETPlus-e-r is the best RFT for the dynamic metrics tested and this is exactly why NETPlus-e-r was selected to be used in the HQUEMU~\cite{hong2012hqemu}.

On top of that, we presented a comprehensive study about several RFTs; we identified the strongest and weakest points of each tested RFT, showing the importance of using RAIn as a tool for the design of any future DTE.
For instance, if one needs to reduce the number of transitions and increase the average time spent in a region, the best RFT would be the Expanded and Relaxed version of NETPlus.
Alternatively, if one needs to create smaller regions with larger completion ration, MRET2 is by far the best tested RFT.

\bibliography{vanderson}

\end{document}